\documentstyle[aps,twocolumn,epsf,amstex]{revtex}

\begin{document}
\draft
\title
{Exclusive photoproduction of hard dijets and   
 magnetic susceptibility of QCD vacuum  \\}
\author{V.~M.~Braun$^1$, S.~Gottwald$^1$, D.~Yu.~Ivanov$^{1,2}$, A.~Sch\"afer$^1$,
        L.~Szymanowski$^{3,4}$ \\}

\address{{}$^1$ Institut f\"ur Theoretische Physik, Universit\"at
          Regensburg, D-93040 Regensburg, Germany }
\address{{}$^2$ Institute of Mathematics, 630090 Novosibirsk, Russia}  
\address{{}$^3$CPhT, \'Ecole Polytechnique, F-91128 Palaiseau, 
    France\thanks{Unit{\'e} mixte C7644 du CNRS}}
\address{{}$^4$ Soltan Institute for Nuclear Studies,
Hoza 69, 00-681 Warsaw, Poland \\}

\date{\today}
\maketitle

\begin{abstract}
We argue that coherent production of hard dijets by linearly 
polarized real photons can provide direct 
evidence for chirality violation in hard processes, the 
first measurement of the magnetic susceptibility of the quark condensate
and the photon distribution amplitude.    
It can also serve as a sensitive probe of the generalized gluon parton 
distribution. Numerical calculations are presented for HERA kinematics.   
\end{abstract}



\narrowtext

The QCD vacuum is a highly complex state. It has a nontrivial particle and 
energy density characterized by quark and gluon condensates and complicated 
dynamical properties that characterize its response to external probes.    
In particular,
consider quarks and antiquarks in the QCD vacuum 
in a constant (electro)magnetic field \cite{IS84}. In the weak field, the induced  
magnetization of the vacuum is proportional to the applied field, the 
quark density $\langle \bar q q\rangle$, the quark electric charge $e_q$ and 
a constant $\chi$ that is called magnetic susceptibility. 
In relativistic notation  
\begin{equation}
   \langle 0|\bar q\sigma_{\alpha\beta} q |0\rangle_F =  
    \,e_q\, \chi\, \langle \bar q q\rangle \,F_{\alpha\beta}
\label{chi2}
\end{equation} 
where $F_{\alpha\beta}$ denotes the electromagnetic field strength.
If the magnetic field 
is varying with a certain frequency, 
the magnetic susceptibility is replaced by 
the corresponding response function which is nothing else but the 
photon distribution function in the infinite momentum frame. 

To be more precise, the wave function of a real photon contains both the perturbative 
chiral-even (CE) contribution of 
the quark-antiquark pair with opposite helicities, and the nonperturbative 
chiral-odd (CO) contribution with quarks having the same helicity and which 
is due to the chiral symmetry breaking. The perturbative CE contribution is 
singular $\sim 1/|{\bf r}|$ at small transverse distances ${\bf r}$ and
well known:
\begin{eqnarray}
\lefteqn{\langle 0|\bar q(0)\gamma_+\frac{1\!\pm\!\gamma_5}{2} q(x)
      |\gamma^{(\lambda)}(q)\rangle 
= \hspace*{2cm}}\label{wfpert}
 \\&=&
 \frac{iN_c e_q }{4\pi^2 {\bf r}^2} q_+\! \int\limits_0^1 \!du\,e^{-iu(qx)}
\left[(e^{(\lambda)}\cdot {\bf r}) (2u-1) 
 \pm i\epsilon_{ik}{\bf r}_i e^{(\lambda)}_k \right],  
\nonumber
\end{eqnarray}    
where $\epsilon_{ik}= \epsilon_{ik+-}$ is the two-dimensional antisymmetric 
tensor in the transverse plane,
$e_u = (2/3)\sqrt{4\pi\alpha_{EM}}$, $e_d = -(1/3)\sqrt{4\pi\alpha_{EM}}$, etc.
 The nonperturbative CO contribution is regular at small transverse separations
(apart from the logarithms) and can be parametrized by the photon distribution 
amplitude $\phi_\gamma(u,\mu)$ \cite{BBK89}
\begin{eqnarray}\label{def3:phi}
\lefteqn{\langle 0 |\bar q(0) \sigma_{\alpha\beta} q(x) 
   | \gamma^{(\lambda)}(q)\rangle = \hspace*{2cm}}
\nonumber\\&=&       
 i \,e_q\, \chi\, \langle \bar q q \rangle
 \left( e^{(\lambda)}_\alpha q_\beta-  e^{(\lambda)}_\beta q_\alpha\right)  
 \int\limits_0^1 \!du\, e^{-iu(qx)}\, \phi_\gamma(u,\mu)\,.
\label{phigamma}
\end{eqnarray}    
Here the normalization is chosen as $\int du\,\phi_\gamma(u) =1$, 
and $u$ stands for the momentum fraction carried by the quark.

The distribution amplitude $\phi_\gamma(u, \mu \ge 1~\mbox{\rm GeV})$ 
is believed to be not far from the asymptotic form
\begin{equation}
   \phi_\gamma^{\rm as}(u) = 6 u (1-u)\,.
\label{phias}
\end{equation} 
The magnetic susceptibility was estimated using the vector dominance 
approximation and QCD sum rules~\cite{chiold,chi}:
\begin{equation}
      \chi \langle \bar q q \rangle \simeq 40-70~\mbox{\rm  MeV}\qquad
       (\mbox{\rm at}~\mu = 1~\mbox{\rm GeV})\,. 
\label{chiestimate}
\end{equation} 
However, any direct experimental evidence on both $\chi$ and $\phi_\gamma(u)$  
is absent. The CO contribution in photoproduction was discussed only once 
in \cite{ht} for the vector meson production at large~$t$.

In this letter we argue that this structure can be studied in experiments for 
the exclusive hard dijet production off nucleons (and nuclei)
\begin{equation}
   \gamma+N \to (\bar q q)+N\,,
\label{dijets}
\end{equation}  
similar to the recent studies of coherent dijet production by incident 
pions by the E791 collaboration \cite{E791}. In particular, we will 
show that perturbative (chirality conserving) and nonperturbative 
(chirality violating) contributions can be separated by the different 
dependence on the longitudinal momentum of the dijets and on the
azimuthal angle. 

The approach of \cite{E791} is different as compared to 
to earlier studies of the dijet photoproduction \cite{H1} 
in that the exclusive dijet final state is 
identified by requiring that the jet transverse momenta are compensated 
to a high accuracy within the diffractive cone and making some additional 
cuts. This approach seems to work for the case of coherent 
dijet production from nuclei by incident pions, and for photoproduction 
the corresponding experimental program  is under way at HERA \cite{Ashery02}.
We assume that separation of the exclusive $q\bar q$ 
dijet final state is feasible.

\begin{figure}[t]
\centerline{\epsfxsize8.0cm\epsffile{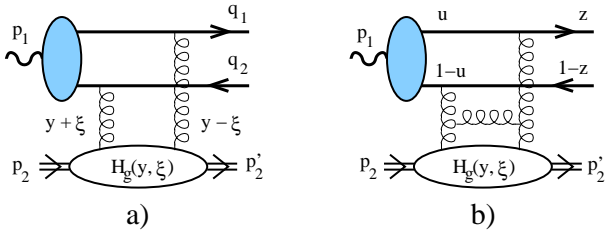}}
\vskip0.2cm
\caption[]{Sample diagrams for the hard dijet photoproduction, see text.  
}
\label{fig:1}
\end{figure}

The kinematics of the process and the notation for the 
momenta is shown in Fig.~\ref{fig:1}.
The Sudakov decomposition of the jet momenta with respect to
the momenta of the incoming particles $p_1$ and $p_2$ reads  
\begin{equation}
q_1=zp_1+\frac{q_{\perp}^2}{zs}p_2+q_{\perp}\,,
q_2=\bar zp_1+\frac{q_{\perp}^2}{\bar zs}p_2-q_{\perp}
\label{Sudakov}
\end{equation}      
so that $z$ is the longitudinal momentum fraction 
and $q_\perp$ the transverse momentum of the quark
jet.
Hereafter 
we use the shorthand notation $\bar u \equiv (1-u)$ for any
longitudinal momentum fraction $u$.
Note that we consider the forward limit, 
when transverse momenta of the jets compensate each other.
In this kinematics the invariant mass of the produced $q\bar q$ pair is
equal to  
$         
M^2={ q_{\perp}^2}/{ z\bar z}
$, 
and the momentum of the outgoing nucleon $p_2^\prime=p_2(1-\xi)/(1+\xi)$, where
$\xi= M^2/(2s-M^2) \simeq M^2/2s$, $s=(p_1+p_2)^2$.         

Since the CE and CO contributions lead to final states with 
different helicity, they do not interfere and the dijet  
cross section is given by the incoherent sum, for the linearly polarized
photon
\begin{eqnarray}
&& 2\pi \frac{d\sigma_{\gamma\to 2\,{\rm jets}}}{d\phi 
d q_\perp^2 dt dz}{\Bigg|}_{t=0} 
\ =\ \sum_q e^2_q \frac{\alpha_{EM} \,\alpha_s^2 \pi^2 
(1+\xi )^2
}{2N_c q_\perp^6}  \nonumber \\
&& \times\left[(1-4z\bar z \cos^2\!{\phi})
|{\cal J}_{CE}|^2+\frac{\pi^2 \alpha_s^2 \chi^2 \langle 
\bar q q \rangle^2 }{N_c^2q^2_\perp}|{\cal J}_{CO}|^2\right],
\label{c.sect.}
\end{eqnarray}
where $\phi$ is the azimuthal angle between the jet direction and 
the photon polarization $(e^{(\lambda)}\cdot q_\perp) \sim \cos\phi$,
${\cal J}_{CE}$ and ${\cal J}_{CE}$ are the CE and CO amplitudes, 
respectively,   
and the prefactors are introduced for future convenience. 
Note that the CE contribution is $\sim 1/q_\perp^6$ \cite{FMS93} and the 
CO contribution is suppressed by one extra 
power of $q_\perp^2$ which follows from twist counting.
The different $\phi$ dependence can be traced to the fact that  
the $q\bar q$ pair is produced in a state with orbital angular momentum 
$L_z=0$ and $L_z=\pm 1$ for the CO and CE contributions, respectively.

The CE contribution originates from the region of large momenta flowing through 
the photon vertex and was considered previously in \cite{NNN,BLW96,LMRT97,GKM98}
in the high energy limit using $k_\perp$ factorization.
We believe that the collinear factorization is more adequate for HERA energies and,
in difference to the dijet production by incident pions \cite{BISS}, expect that  
it is valid for the CE amplitude to all orders in perturbation theory.
Hence the amplitude ${\cal J}_{CE}$ is given by the convolution integral of 
the coefficient function and the generalized parton distribution, 
cf.~\cite{BISS}. For high energies the gluon contribution is dominant.
To leading order (LO) in the strong coupling $\alpha_s = \alpha_s(q_\perp)$
the amplitude is given by the sum of Feynman diagrams of the type 
shown in Fig.~\ref{fig:1}a with all possible attachments of the gluons.
The answer reads
\begin{equation}          
{\cal J}^{LO}_{CE}=-\frac{1}{\pi}
\int\limits^1_{-1}\!
dy\, {\cal H}_{g}(y,\xi)
\frac{\xi}{(y-\xi+i\epsilon)^2}\,,  
\label{2pole}
\end{equation}
where ${\cal H}_{g}(y,\xi) = {\cal H}_{g}(-y,\xi)$ is the generalized 
gluon distribution in the 
symmetric notation \cite{Ji96}, 
$y+\xi$ and $y-\xi$
are the $t$-channel gluon momentum fractions with respect to $(p_2+p_2')/2$.
The full calculation of the NLO contribution goes beyond the tasks of this 
letter. For the reasons explained below it is necessary, however, to 
include the leading NLO contribution at large energies (enhanced by $\log\xi$)
which corresponds to an additional gluon exchange between the    
$t$-channel gluons, see Fig.~\ref{fig:1}b. Including this contribution, 
the imaginary part of the amplitude equals
\begin{equation}
Im {\cal J}_{CE}= \xi {\cal H}_{g}^\prime (\xi,\xi)
+\frac{\alpha_s N_c}{\pi}\int^1_\xi\frac{dy }{y+\xi}
{\cal H}_{g} (y,\xi)\,,
\label{even}
\end{equation}
where ${\cal H}_{g}^\prime (\xi,\xi)=d{\cal H}_{g} (y,\xi)/dy|_{y=\xi}$.
The real part is smaller and can be neglected in the first approximation.

For high energies alias $y\to 0$, \,  
${\cal H}_{g}(y,\xi)\sim y^{-\Delta}$, where  in perturbation theory $\Delta \sim 
\alpha_s\log{s/q^2_\perp}$ has to be treated as
a small parameter. Therefore, despite the fact that the two terms in
(\ref{even}) appear in different orders in the collinear expansion,
they are of the same order as far as the counting of energy logarithms
is concerned. This feature is specific for real photons and can be 
traced to the fact that the LO amplitude in (\ref{2pole}) only contains 
a (rather unusual) double-pole, but no single poles (c.f. (\ref{chodd})). 
Since ${\cal H}_{g} (y,\xi)\sim 
G(y)$ at $y\gg \xi$, and as the factor 
$\alpha_s N_c/(\pi y)$ is nothing but the low-$y$ limit of the 
DGLAP gluon splitting function,  the integral in Eq.~(\ref{even}) can be 
identified to logarithmic accuracy with the unintegrated gluon distribution 
$f(\xi,q^2)=\partial G(\xi,q^2)/\partial \ln q^2$. This contribution corresponds to the
one considered in \cite{NNN,BLW96,LMRT97} in the $k_\perp$ factorization approach.
The first contribution in (\ref{even}) is analogous to Eq.~(42) in \cite{GKM98}.  

For the nonperturbative CO contribution 
the large momenta are not allowed in the 
photon vertex and the factorization formula contains a 
convolution with the photon distribution amplitude. 
In this case an additional hard gluon exchange is mandatory and the 
diagram in Fig.~\ref{fig:1}b presents one example of the existing 31 LO contributions,
cf. Fig.~11 in Ref.~\cite{BISS}. The calculation of this contribution 
is similar to the case of pion diffraction dijet production considered in much
detail in \cite{BISS}. Here we only present the final result 
($C_F=(N^2_c-1)/(2N_c)$):
\begin{eqnarray}
&&{} 
{\cal J}_{CO}= -\frac{1}{\pi }\int\limits^1_{-1} dy
\int\limits^1_0 du {\cal H}_g(y,\xi)\frac{\phi_\gamma (u)}{u\bar u}
\bigg\{  \nonumber \\
&&{}
C_F
\left(\frac{2\xi}{(y-\xi +i\epsilon )^2}
-\frac{1}{y-\xi +i\epsilon }
\right) + \left(C_F\left(\frac{z\bar z}{u\bar u}+1 \right) 
 \right.
\nonumber \\
&&{}
\left.
+ \frac{1}{2N_c}\left(\frac{z}{u}+\frac{\bar z}{\bar u}
\right)\right)
\frac{z\bar u+u\bar z}{y(z-u)-\xi (z\bar u+u \bar z)+i\epsilon}
 \nonumber \\
&&{}
\left.
-\left(
C_F \frac{z\bar z}{u\bar u}
+\frac{1}{2N_c}\left(
\frac{z}{u}+\frac{\bar z}{\bar u}
\right)\right)
\frac{1}{(y-\xi-i\epsilon)}
\right\}.  
\label{chodd}
\end{eqnarray}
Integration over the quark longitudinal momentum fraction contains a logarithmic 
divergence at the end points $u\to 0,1$ which signals that the collinear factorization
is violated. The divergent contribution is purely imaginary and reads
\begin{equation}
{\cal J}_{CO}^{IR}=2i{\cal H}_g(\xi,\xi)
\left(N_c z\bar z+\frac{z^2+\bar z^2}{2N_c}\right)
\int^1_{u_{\rm min}}\!\!\frac{du}{u^2}\, \phi_\gamma (u)\,, 
\label{e.p.}
\end{equation}
where we have introduced an infrared cutoff $u_{\rm  min} = \mu^2_{\rm IR}/q_\perp^2$.
In numerical calculations we use $\mu_{\rm IR}=500$~MeV.
The origin of the factorization breaking is that both final and initial state 
interactions are present and lead to pinching of the integration 
contour in the 
so-called Glauber region, see \cite{BISS} for a detailed discussion. 
In the present context violation of factorization is probably 
not surprising since the CO contribution is suppressed by a power of $q_\perp^2$ 
compared to the leading twist. 

Another important integration region for the quark momentum fraction in (\ref{chodd}) is 
$\xi \ll |u-z| \ll 1$ which produces a logarithmic enhancement in energy:
\begin{equation} 
{\cal J}_{CO}^{DLA}=4iN_c\phi_\gamma (z)\int^1_\xi\frac{dy}{y+\xi}\,
{\cal H}_g(y,\xi)\,.
\label{h.e.}
\end{equation}   
Hence in the double-logarithmic approximation collinear 
factorization is valid for the CO contribution as well. 

Assuming that the photon distribution amplitude 
$\phi_\gamma (z,\mu=q_\perp)$ is close to the asymptotic form (\ref{phias}),
we obtain ${\cal J}_{CO}\sim z(1-z)$ for both integration regions, up to small 
corrections. The IR divergence in (\ref{e.p.}) does not have, therefore, 
any significant effect on the jet distribution but mainly influences the 
normalization.  

\vskip0.5cm

\begin{figure}[t]
\centerline{\epsfxsize6.5cm\epsffile{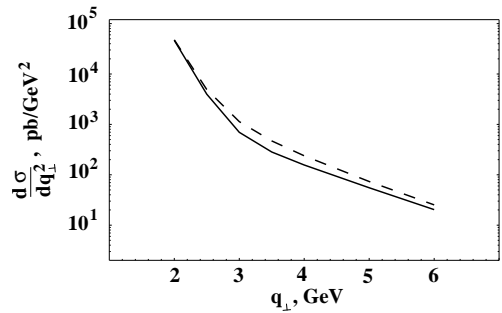}}
\vskip0.2cm
\caption[]{The cross section $d\sigma_{\gamma+N\to 2\,{\rm jets}+N}/ 
d q_\perp^2$, see text.
The dashed curve corresponds to the calculation with the lower limit in the 
integral in (\ref{even}) changed to $2\xi$ and a different IR-cutoff 
$\mu_{\rm IR}=350$~MeV.    
}
\label{q2dependence}
\end{figure}

In the numerical calculation presented below we have taken into account 
the full result in (\ref{even}) and the imaginary part of the CO contribution 
in (\ref{chodd}). Calculations are done for HERA kinematics, $s=10000$~GeV$^2$ and 
the value $\chi \langle \bar q q\rangle = 50$~MeV, cf. (\ref{chiestimate}).
We use the parametrization of the generalized gluon   
distribution by Freund and McDermott \cite{GPDs} that is based on the  
MRST2001 leading-order forward distribution \cite{MRST2001}. The transverse momentum 
dependence of the cross section integrated over $\phi$, $z$ and $t$~\cite{t-dependence} 
is shown in Fig.~\ref{q2dependence}.
\noindent
For $q_\perp > 4$~GeV the cross section is dominated by the perturbative CE contribution.
The nonperturbative CO contribution at $q_\perp\sim 4-6$~GeV is of the order of
\begin{equation}
 \frac{d\sigma_{CO}/dq_\perp^2}{d\sigma_{CE}/dq_\perp^2} \simeq
   (7\pm 2~\mbox{\rm GeV})^2\cdot \frac{\alpha_s(q_\perp)^2}{q_\perp^2}\,
   \left(\frac{\chi\langle \bar q q\rangle}{50~\mbox{\small \rm MeV}}\right)^2.
\label{COscale}
\end{equation}
Note a large mass scale which is due to the kinematical enhancement of the CO 
amplitude by a large factor $4\pi^2$, compare (\ref{wfpert}) and (\ref{phigamma}).
On top of this, the two contributions in (\ref{even}) have opposite sign and 
tend to cancel. For large transverse momenta the first term dominates,
which is natural since it is leading order in the collinear expansion. However,
progressing towards lower $q_\perp$ one effectively goes over to smaller values 
of Bjorken $x\sim 2 \xi$ so that the second contribution eventually becomes larger
than the first one and the imaginary part of CE amplitude changes sign. 
This explains an abrupt change of the slope of the 
cross section at $q_\perp\sim 3$~GeV. For smaller transverse momenta the dijet cross
section is dominated by the CO contribution. Because of the cancellations in the CE 
contribution at $q_\perp\sim 2.5-3.5$~GeV the complete NLO calculation is 
required to make the predictions in this region fully quantitative. 
As a crude estimate of the uncertainties  we show as dashed curve 
in Fig.~\ref{q2dependence} the results if  the lower limit in the 
integral in (\ref{even}) is changed to $2\xi$ and with a different IR-cutoff 
$\mu_{\rm IR}=350$~MeV.  
  
\begin{figure}[t]
\centerline{\epsfxsize5.0cm\epsffile{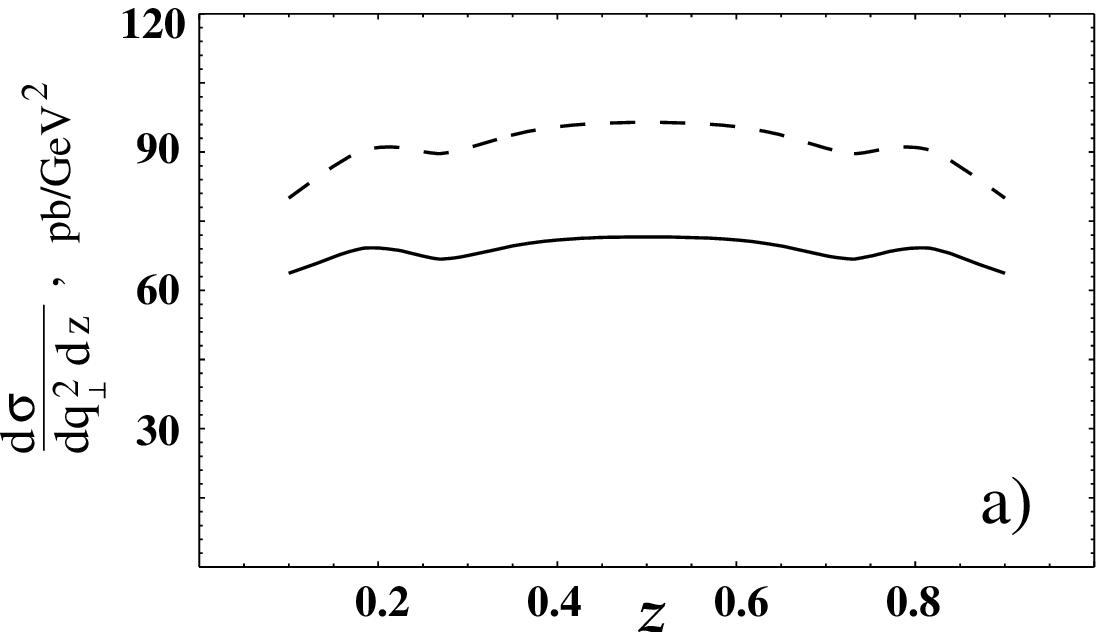}}
\centerline{\epsfxsize4.9cm\epsffile{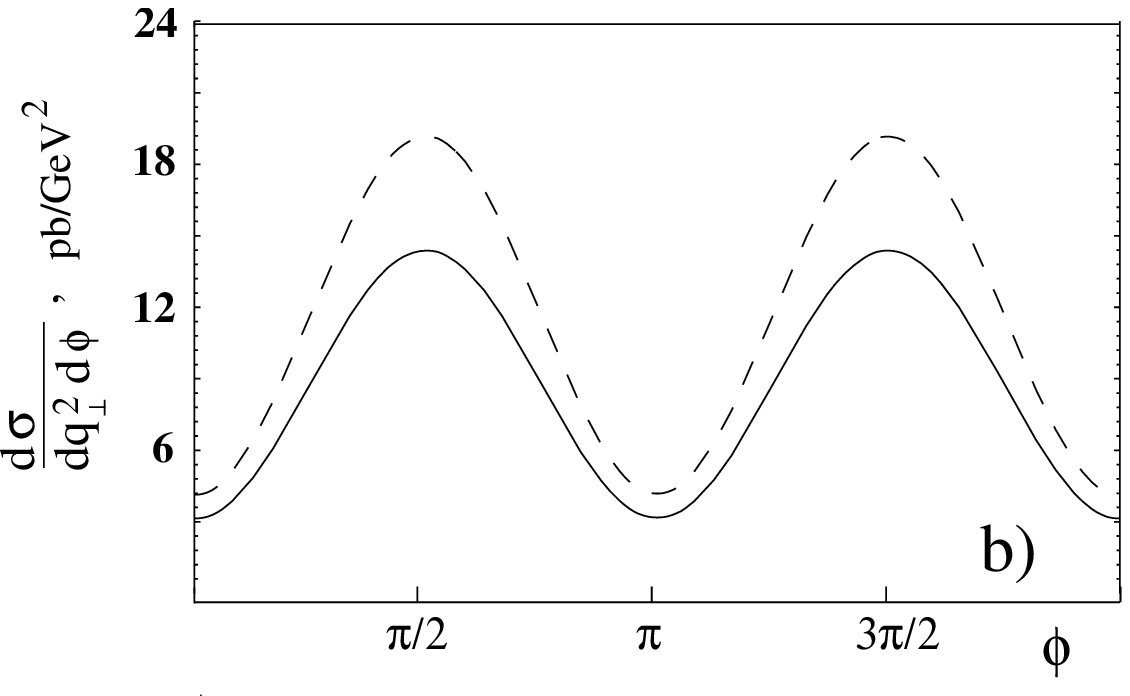}}
\vskip0.2cm
\caption[]{The differential cross section 
$d\sigma_{\gamma+N\to 2\,{\rm jets}+N}/d q_\perp^2 dz$ (a) and ,
$d\sigma_{\gamma+N\to 2\,{\rm jets}+N}/d q_\perp^2 d\phi$ (b)
for jet transverse momentum $q_\perp=5$~GeV.  
Identification of the curves is the same as in Fig.~\ref{q2dependence}. 
}
\label{5GeV}
\end{figure}

\begin{figure}[t]
\centerline{\epsfxsize5.0cm\epsffile{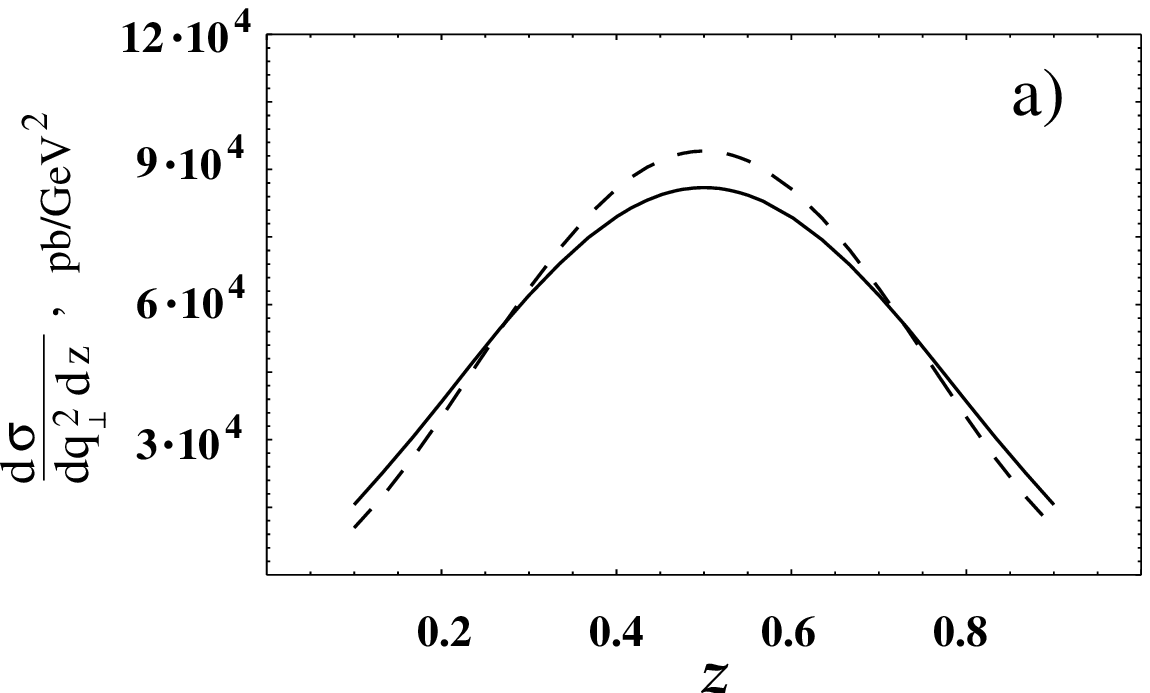}}
\centerline{\epsfxsize4.9cm\epsffile{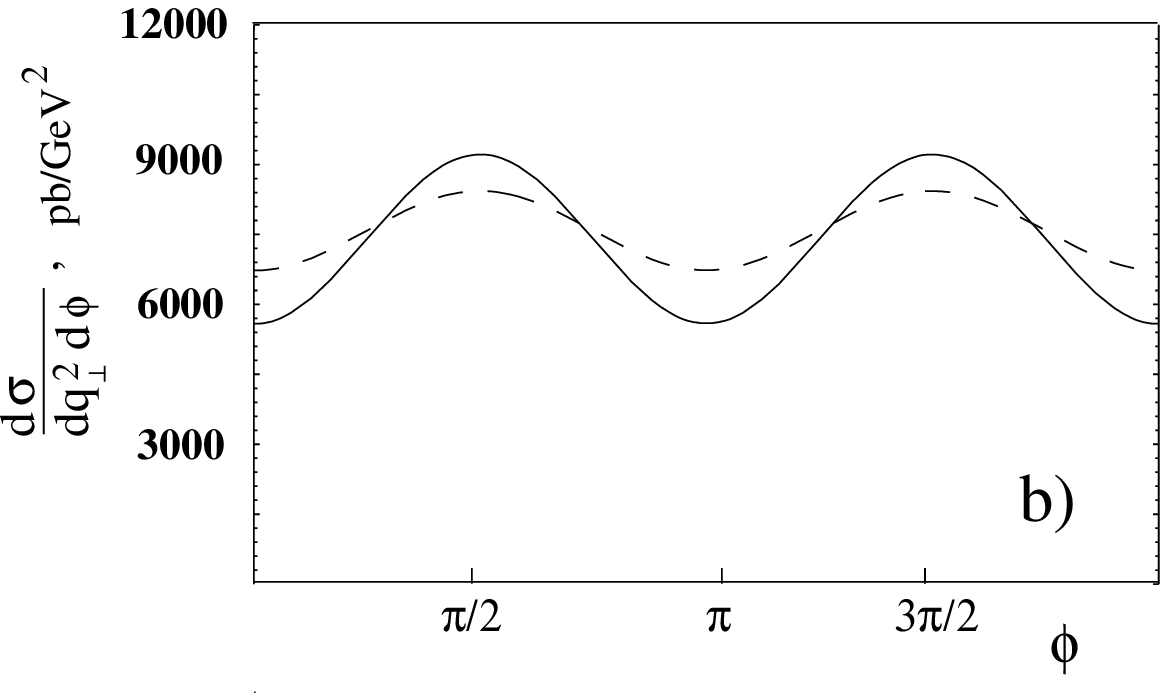}}
\vskip0.2cm
\caption[]{
Same as in Fig.~\ref{5GeV} but for $q_\perp=2$~GeV.  
}
\label{2GeV}
\end{figure}

The transition between the two different regimes is seen very clearly from 
the dependence of the cross section on the dijet longitudinal momentum fraction and 
the azimuthal angle, compare Fig.~\ref{5GeV} and  Fig.~\ref{2GeV}.
At $q_\perp = 5$~GeV the (parton level) $z$-distribution is almost flat, while 
the $\phi$ distribution is almost purely $\sim 1-\cos^2\phi$. In contrast to this,
at $q_\perp = 2$~GeV the $z$-distribution is comparable with  $\sim z^2(1-z)^2$
while the $\phi$-distribution is flat.   

To avoid misunderstanding, we repeat that all calculations in this paper are done 
for the fixed value $\chi \langle \bar q q\rangle = 50$~MeV and one particular 
model \cite{GPDs} of the generalized gluon distribution. The related uncertainties 
are not included. In fact, coherent photoproduction of dijets may present the 
best opportunity to constrain both of them. 
\noindent

To conclude, we summarize our main points. 
In this letter we argue that studies of exclusive photoproduction
of dijets with large transverse momenta can yield important 
information on the photon structure at small distances.
Our main result is that the nonperturbative CO contribution is large in the region
of intermediate $q_\perp\sim 2-4$~GeV and can be clearly separated from the 
perturbative contribution by a different $z$- and $\phi$-dependence. 
Observation of the CO contribution would be the first clear
evidence for the chirality violation in hard processes and also provide the 
first direct measurement of the magnetic susceptibility of the quark condensate.    
On the other hand, the dijet cross section for large $q_\perp$ can serve to
constrain the generalized gluon distribution.

On the theoretical side, we deviate from previous studies of the dijet 
production by consistently applying the collinear factorization
in terms of generalized parton distributions. For the nonperturbative CO contributions
the collinear factorization is, strictly speaking, broken. However, the sensitivity 
to the IR cutoff is relatively weak and can formally be eliminated by taking into 
account Sudakov-type corrections in the modified collinear factorization framework.
We think that this technique is potentially more accurate and the results can be improved 
systematically by the calculations of higher-order corrections. In particular, the 
complete NLO calculation of the perturbative CE contribution would be very welcome  because
of cancellations that are discussed in the text.

\vskip0.5cm

\paragraph*{Acknowledgements.}
D.I. thanks the Alexander von Humboldt Foundation for the financial 
support. The work by S.G. was funded by the DFG grant 920770.
L.S. is grateful to B.~Pire for discussions and hospitality at
Ecole Polytechnique.


\begin{references}
\bibitem{t-dependence}
For the $t$-integration we assume  
$d\sigma/dt \sim \exp(bt)$ with the slope $b=6$~GeV$^2$.
\end{references}
\end{document}